\documentclass[preprint,12pt]{elsarticle}

\usepackage{amssymb}

\journal{Chemical Physics Letters}

\usepackage{graphicx}
\usepackage{dcolumn}
\usepackage{bm}
\usepackage{amsmath}
\usepackage{tabularx}
\begin{document}

\begin{frontmatter}



\title{Metastable structure of photoexcited WO$_{3}$ determined by the pump--probe extended X-ray absorption fine structure spectroscopy and constrained thorough search analysis}


\author[label1]{Daiki Kido}
\author[label2]{Yohei Uemura}
\author[label2]{Yuki Wakisaka}
\author[label2]{Akihiro Koide}
\author[label1]{Hiromitsu Uehara}
\author[label3]{Yasuhiro Niwa}
\author[label3]{Shunsuke Nozawa}
\author[label3]{Kohei Ichiyanagi}
\author[label3]{Ryo Fukaya}
\author[label3]{Shin-ichi Adachi}
\author[label4]{Tokushi Sato}
\author[label5]{Harry Jenkins}
\author[label2]{Toshihiko Yokoyama}
\author[label1]{Satoru Takakusagi}
\author[label1]{Jun-ya Hasegawa}
\author[label1]{Kiyotaka Asakura*}

\affiliation[label1]{organization={Institute for Catalysis Hokkaido University},
            addressline={Kita21, Nishi10, Kita-ku}, 
            city={Sapporo},
            postcode={001-0021}, 
            state={Hokkaido},
            country={Japan}}
            
\affiliation[label2]{organization={ Institue for Molecular Scinence},
            addressline={38 Nishigo-Naka,Myodaiji}, 
            city={Okazaki},
            postcode={444-8585}, 
            state={Aichi},
            country={Japan}}

\affiliation[label3]{organization={Photon Factory, Institute for Materials Structure Science},
            addressline={1-1 Oho,}, 
            city={Tsukuba},
            postcode={305-0801}, 
            state={Ibaraki},
            country={Japan}}

\affiliation[label4]{organization={European XFEL GmbH},
            addressline={4}, 
            city={Holzkoppel},
            postcode={22869}, 
            state={Schenefeld},
            country={Germany}}

\affiliation[label5]{organization={Cardiff Catalysis Institute, Cardiff University},
            addressline={}, 
            city={Cardiff},
            postcode={CF10 3AT}, 
            state={Wales},
            country={UK}}

\begin{abstract}
We have determined the local structure of photoexcited metastable WO$_3$ created 150 ps after  400 nm laser irradiation by the pulse pump--probe L$_3$-edge extended X-ray absorption fine structure spectroscopy and the constrained thorough search analysis.  We have found a highly distorted octahedral local structure with one of the shortest  W=O bonds  being further shortened to 1.66 \AA\  while the other five bonds were elongated even though theoretical calculations predicted the reverse change. We discuss this contradiction and propose a possible structure for the metastable state.
\end{abstract}

\begin{graphicalabstract}
\includegraphics[scale=0.9]{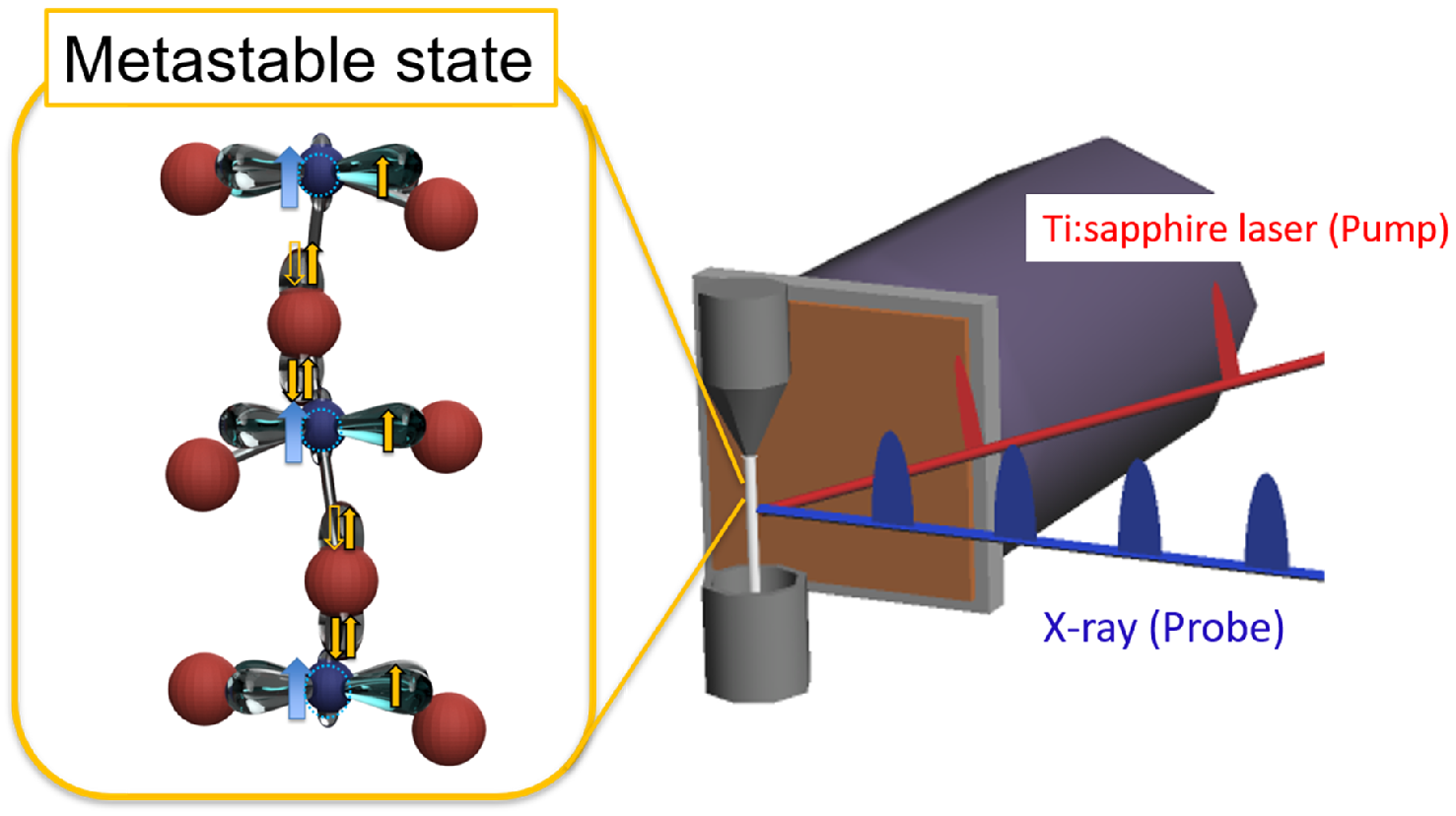}
\end{graphicalabstract}

\begin{highlights}
\item Research highlight 1

The structure of WO$_3$ in the matastable state was determined by the pump-probe EXAFS measurement and constrained thorough search analysis

\item Research highlight 2

The shortest  W=O bonds  being further shortened to 1.66 \AA \ 

\item Research highlight 3

The well-ordered distorted triplet cluster formation was proposed
\end{highlights}


\end{frontmatter}


\section{Introduction}
\label{sec:introduction}


Semiconductor oxides are used as photocatalysts to harvest sunlight to split water into H$_2$ and O$_2$.\cite{Strunk2021}
Electrons in the valence band, which is mainly composed of O 2$p$ orbitals, are excited by photoirradiation to the conduction band, followed by stabilization of the photoelectrons to form the metastable state (MS), which is often assumed to be associated with surface or bulk defect sites.\cite{Strunk2021} 
The local structure and electronic state of the MS need to be elucidate to understand the photocatalysis of water splitting and to improve the performance of water-splitting photocatalysts.

A pump--probe (PP) X-ray absorption fine structure (XAFS) investigation of the photocatalysts using pulsed X-rays emitted from an X-ray free electron laser (XFEL) and synchrotron radiation (SR) has revealed element-specific local electronic and geometrical structures, thereby providing  new scientific knowledge about photocatalysis that cannot be obtained by other techniques.\cite{RN7738} 
WO$_3$ is used as a Z-scheme water-splitting photocatalyst that is sensitive to visible light, which is the main component of sunlight.\cite{abe2010} 
We have acquired PP X-ray absorption near edge structure (PP-XANES) spectra of WO$_3$ after pulsed laser irradiation and have found that the MS was created within 150 ps, followed by relaxation to the ground state (GS) with a time constant of 2 ns.\cite{RN257,RN25,RN2588,RN7257} 
We proposed that the photoelectrons were trapped at distorted stoichiometric W sites in the crystal lattice, similar to the trapping of polarons, instead of at surface or bulk defects.\cite{RN2588}
The formation time for the MS, 150 ps, was longer than that for the other oxides such as TiO$_2$ and Fe$_2$O$_3$, consistent with previously-reported optical measurement results\cite{RN2269}.
It is desirable to acquire the PP extended X-ray absorption fine structure (PP-EXAFS) spectrum to confirm the structure change because EXAFS is sensitive to local structural features such as bond lengths.\cite{RN1856}
However, PP-EXAFS spectral measurements are much more difficult than PP-XANES measurements \cite{RN2588} because of the lower PP-EXAFS signal intensity.  In addition compounds with a complex structure, such as WO$_3$, require a multishell fitting which is inherently challenging.  
We addressed these two difficulties by taking the difference spectra between before and after photoabsorption to obtain the EXAFS spectrum of MS WO$_3$. 
The difference spectra were  accumulated and then analyzed using the constrained thorough search (CTS) method to elucidate the bond lengths in MS WO$_3$.\cite{RN7005,RN8640} 

In the present paper, we report the PP-EXAFS results for the structure of the MS WO$_3$ state to clarify the local structure change and to attempt to explain the long formation time for MS WO$_3$.

\section{Experimental}
\label{sec:experimental}
\subsection{Materials and methods}
The WO$_3$ was purchased from Wako Chemicals, and 50--200 nm WO$_3$ particles were dispersed in ultrapure water.  
The concentration of WO$_3$ was 2.5 mmol L$^{-1}$.
The WO$_3$ jet was supplied to the cross point of the laser used for photoexcitation (Ti-sapphire with a 1 ps pulse width, 945 Hz repetition rate, and 400 nm wavelength) and the X-ray beam emitted from the Photon Factory Advanced Ring (PF-AR, 6.5 GeV, 60 mA, single bunch operation with pulse width of 100 ps, and pulse interval of 1.26 $\mu$s).
The fluorescence X-rays were monitored using a scintillation counter with a Cu filter to attenuate the elastic X-rays.
Details of the experimental setup are available elsewhere.\cite{RN257}   
\subsection{Analysis}
The difference spectrum before and after the photoirradiation, $\Delta \mu(=\mu_{API}-\mu_{BPI}) $, was accumulated, thereby reducing the possible fluctuations. 
The $\mu_{API}$ and $\mu_{BPI}$ are the spectra recorded after and before the photoirradiation, respectively.
The spectrum of the MS WO$_3$, $\mu_{MS}$, was obtained by assuming that $\mu_{API}$ was a linear combination of the spectrum of the ground state (GS), $\mu_{GS}$, and $\mu_{MS}$.  
${\mu}_{API}={{\alpha\mu}_{MS}+{\left(1-\alpha\right)\mu}_{GS}}$, where $\alpha$, the fraction of MS WO$_3$, was obtained from the XANES spectrum, as shown in Fig. \ref{fig:XANES} in the supporting information (\ref{sec:XANES}). 
${\mu}_{MS}$ was processed and analyzed using the EXAFS analysis programs REX\cite{RN3616} and LARCH \cite{RN6645}, as described in \ref{sec:Analysis}.
The Fourier transformation ranges of $\Delta k$ and $\Delta r$ were 5 ($=8-3$) \AA$^{-1}$  and 1 ($=2-1$) \AA \ , respectively, so that the amount of information, $M$, was approximately 5 ($M=2 \Delta k \cdot \Delta r)/\pi + 2)$\cite{PhysRevB.48.9825}. 
The inversely Fourier transformed data were first analyzed by a curve-fitting method using the LARCH package.\cite{RN6645}
Further  details of the structure, including bond lengths, were derived from the EXAFS spectra using the CTS method, as explained in \ref{sec:CTS}.\cite{RN7005,RN8640,RN7246}

\section{Results}
\label{sec:Results}
\begin{figure}
 \centering 
  \includegraphics[width=120mm,keepaspectratio,clip]{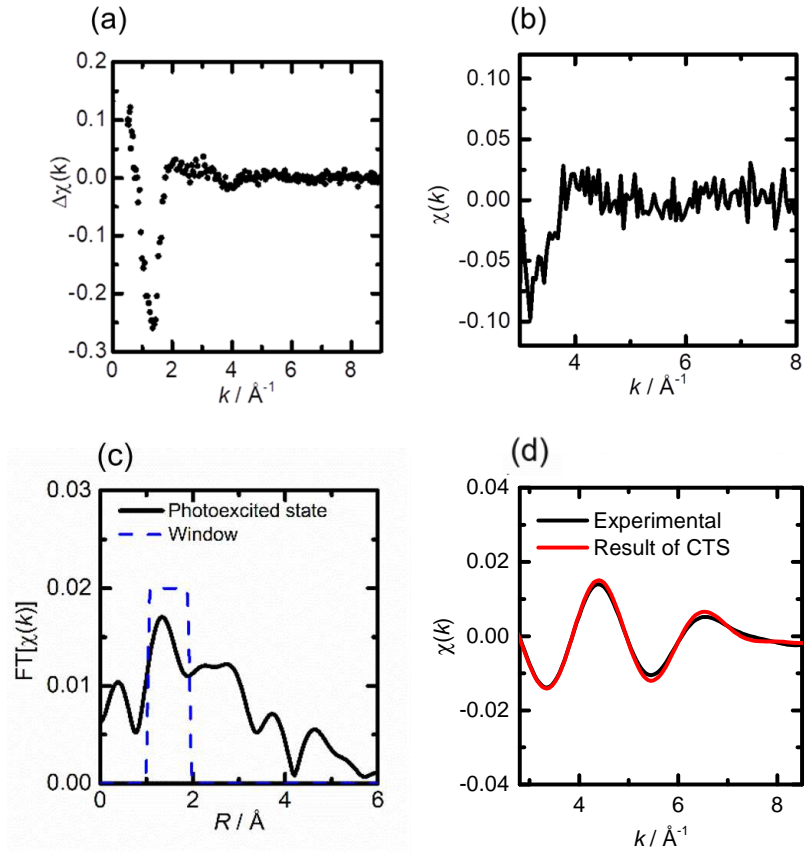}
  \caption{(a)$\Delta\chi\left(k\right)$ for W L$_3$-edge EXAFS spectra between before and 150 ps after the irradiation. (b) $\chi_{MS}\left(k\right)$  for the MS WO$_3$.  (c) The Fourier transform of $\chi_{MS}\left(k\right)$. The broken line shows the range of Fourier filtering. (d) Inversely Fourier transformed data(black line) for the MS WO$_3$ with the results of CTS(red line).  The enlarged one is available in Fig. \ref{fig:TSFITTING}}
  \label{chi_1}
\end{figure}

Fig. \ref{chi_1} shows the EXAFS oscillation of the MS WO$_3$, $\chi(k)_{MS}$, and its Fourier transform.
The main peak at 1.4 \AA \  was inversely Fourier transformed to the $k$-space, and the data were analyzed by one-shell curve fitting. 
The one-shell curve fitting results are shown in Fig. \ref{fig:CF} and Table \ref{tab:CF1} in \ref{sec:RCF}. 
We conducted the analysis using two methods.
In the first method, the Debye--Waller (DW) factor was adjusted freely; in the second method, the DW factor was constrained as a function of the bond length given by Eq.\ref{eq:DWANDR} in \ref{sec:CTS}.
Both results indicated that the bond length contracted in the MS WO$_3$; however, details of the structure were unclear.

We carried out the CTS analysis based on the uniform prior probability principle.\cite{RN8640}   
We assumed the followings:
\begin{enumerate}
\item MS WO$_3$ has uniformly distorted octahedra, similar to GS WO$_3$.  
\item  DW factors are functions of the bond length in Eq.\ref{eq:DWANDR} in \ref{sec:CTS}.\cite{RN1924} A greater length corresponds to a larger DW factor.
\item EXAFS spectroscopy is sensitive to short bonds because of the \\
$\exp (-2r/\lambda -2 \sigma^2 k^2 ) / r^2 $ term. 
\end{enumerate}
The structural parameters were limited to the value of $M$ ($\approx$5).  
The GS WO$_3$ has a monoclinic structure in which two distorted octahedral sites are present, as shown in Table \ref{tab:CFR}.\cite{RN8155}
The bond lengths in GS WO$_3$ have 12 different values.  
Two sites have a similar local structure; the corresponding W--O distances are therefore denoted as $r'''_1$, $r'''_2$, $r'''_3$, $r'''_4$, $r'''_5$, and $r'''_6$ in Table \ref{tab:CFR} in \ref{sec:CTS}. 
We intended to demonstrate that $r'''_1 \ne r'''_2$  in the MS WO$_3$, as predicted from the W L$_1$-edge XANES results.\cite{RN2588} 
According to assumption (3) (i.e., that shorter bonds ($r'''_1$, $r'''_2$) have greater contributions to the EXAFS spectra), we assumed that $r'''_1$ and $r'''_2$ were independently optimized and we constrained $r'''_3 = r'''_4 = r_3$ and $r'''_5 = r'''_6=r_4$, as shown in Table \ref{tab:CFR}. The CTS was carried out for the four parameters $r_1(=r'''_1)$, $r_2(=r'''_2)$, $r_3$, and $r_4$ with the corresponding coordination numbers, $N_1=N_2=1$ and $N_3=N_4=2$, as shown in Table \ref{tab:tableS1} and Table \ref{tab:CFR} in \ref{sec:CTS}.  
Table \ref{tab:cts} shows the CTS results that reproduced the crystal data given in Table \ref{tab:CFR} in \ref{sec:CTS} for the GS WO$_3$. 
Note that $r_1 $ and $r_2$ are approximately the same as shown in Table \ref{tab:cts} even if the two parameters are independently changed.
Fig. \ref{fig:histoGS} shows the occurrence histograms.
We found that the histograms for $r_1$ and $r_2$ have similar distributions and the same peak positions.
Consequently, for the GS WO$_3$, we conclude that $r_1 $ and $r_2 $ agreed with each other within the error.  

\begin{figure}

 \centering 
    \includegraphics[width=80 mm]{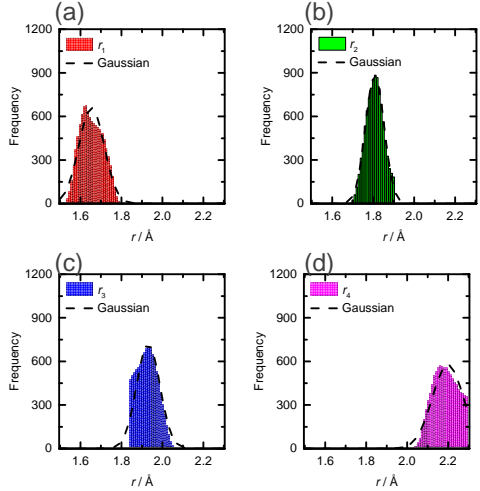}
    \caption{(a)--(d) Distribution of bond lengths that satisfy R-factor $<$ 0.10 for the structure parameters for the MS WO$_3$.  Dashed lines show the fitting results obtained using a Gaussian function. }
    \label{fig:HistoMS}
\end{figure}

Table \ref{tab:cts} shows the CTS results for the MS WO$_3$.  
Fig. \ref{fig:TSFITTING} shows a comparison between the Fourier filtered $\chi_{MS}(k) $ and the calculated one. 
The observed data were well reproduced by the parameters determined by the CTS method. 
The R-factor(eq. \ref{eq:RF}) was 0.02. 
In the case of the MS WO$_3$, we found that $r_1$ and $r_2$ were not equal and that $r_1$ was 0.11 \AA  \ shorter than that for the GS WO$_3$. 
The histograms of $r_1$ and $r_2$ show different peak positions (Fig. \ref{fig:HistoMS}).  
Bonds other than $r_1$ in the MS WO$_3$ were  longer than the corresponding bonds in the GS WO$_3$, in good agreement with previously reported WO$_3$ L$_1$ XANES results.\cite{RN2588}  
The photoexcited MS WO$_3$ exhibited a further distorted structure. 

\begin{table}[htbp]
    \caption{Results of constrained thorough search analysis for the W L$_3$-edge EXAFS spectra of GS and MS WO$_3$.    "Crystal" indicates crystallographic data.\cite{RN8155}  }
        \label{tab:cts}
    \centering
    \begin{tabular}{c c c c c  c}
    \hline
        sample &$r_1$ & $r_2$ & $r_3$ &$r_4$ & \begin{tabular}{c}
R-\\factor \end{tabular}\\
         \hline
GS &1.77$\pm 0.08 $&1.77$\pm 0.08$ &1.91$\pm 0.04 $ &2.17$\pm 0.09$ &0.049\\
         Crystal & (1.75 & 1.77 & 1.90 & 2.13 )&\\
         \hline
         MS
        & 1.66$\pm 0.06$ &1.81$\pm 0.04$ &1.93$\pm 0.06$ &2.19$\pm 0.07$ & 0.012\\
         \hline
    \end{tabular}
\end{table}

\section{Discussion}
\label{sec:Discussion} 
The CTS analysis showed that the MS WO$_3$ exhibited a further distorted structure, where the shortest W=O bond was further shortened from 1.77  \AA \ to 1.66  \AA. \cite{RN2588}
This structure change was interpreted as follows in a previous paper.\cite{RN2588} 
The photoelectron excited by the pulsed laser would occupy the d$_{xy}$ orbital of W, which is on a plane perpendicular to the shortest W=O bond.  The orbital is located at the bottom of the conduction band in energy scale, forming a $\pi^*$ antibonding orbital with 2$p_x$ and 2$p_y$ for oxygen atoms in the $xy$ plane.  
Thus, the W--O bonds in the $xy$ plane were elongated, pushing up the W atom in the direction of the shortest W=O bond.\cite{RN2588}   
As described later and in \ref{sec:MO} and in \ref{sec:qmmm}, this previous interpretation is not consistent with the results of
the density functional theory (DFT) and quantum mechanics/molecular mechanics (QM/MM) calculations.  This discrepancy will be discussed below.

In the literature, the MS WO$_3$ was assumed to be located in bulk or surface defect sites.\cite{RN1799,RN8356, RN2222} 
However, we disproved these possibilities because the amount of MS states is much larger than the amount of defects.  
Yamakata {\it et al.} recently reported that introducing defects by reducing WO$_3$ decreased the lifetime of the MS WO$_3$.\cite{RN8614} 
Surface defects might be less likely because the surrounding environment did not affect the lifetime of the MS WO$_3$, as shown in \ref{sec:react}.
We proposed the MS WO$_3$ should be located at the normal lattice sites in the bulk stoichiometric WO$_3$ and be distorted from the structure of the GS WO$_3$.
Such distortion would stabilize the MS WO$_3$, similar to the stabilization of polarons. 
\begin{figure}[htbp]
 \centering 
  \includegraphics[width=\linewidth,keepaspectratio,clip]{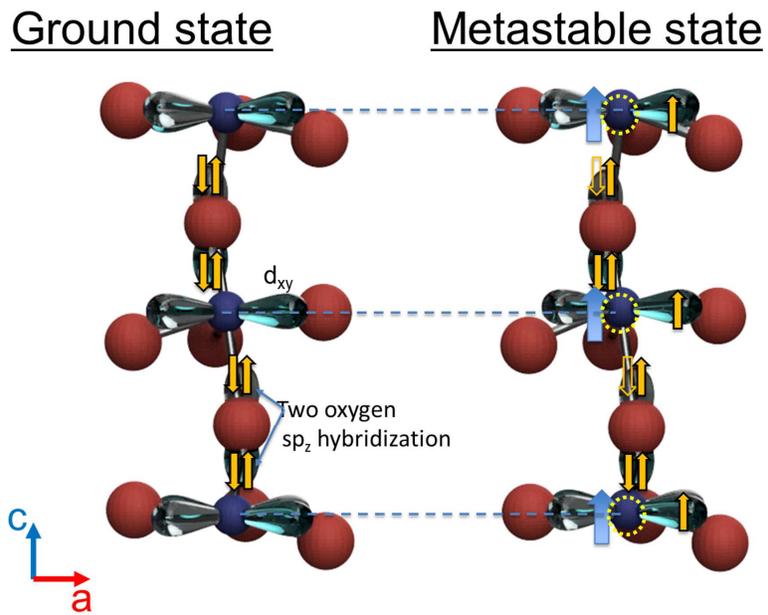}
  \caption{The local structure change in the GS WO$_3$ (left panel) and the MS WO$_3$ ( right panel).   Large red and small blue circles represent O and W atoms, respectively.  yellow broken circles indicate the position of W in GS. Thick blue arrows show the movement of W atoms in the MS WO$_3$.  W atoms along the $c$-axis move in the direction corresponding to W=O contraction. Orange filled and empty arrows in the orbitals correspond to electrons with spin and holes, respectively.  Grey filled circles indicate holes.   The unpaired electron spins are all parallel in the polaron cluster.}
  \label{fig:polaron}
\end{figure}

However, we still faced two problems.
First, the DFT and QM/MM calculations for the MS WO$_3$ local structure indicated that the less-distorted structure was stable, with the shortest W=O bond being elongated as shown in Table \ref{tab:GAUSSIAN} in \ref{sec:MO}. 
Second, it was unclear why the formation of the MS WO$_3$ structure took 150 ps.  
The vibration frequencies of W--O are on the order of 100 cm$^{-1}$ or 3 THz or 0.3 ps. 
Fe$_2$O$_3$ hematite and CuWO$_4$ took less than 1 ps to produce a polaronic MS.\cite{RN7556,RN8560}   
Thus, the simple isolated distorted MS structure should form much faster.  
According to the L$_3$ edge XANES result, the W$^{5+}$ was formed and was gradually changed to the MS structure.\cite{RN25}   
Such a long time scale might be related to the collective structure change. 
If the single distorted structure of the MS WO$_3$ was located in the GS WO$_3$, it should have a greater distortion energy between the surrounding GS WO$_3$. 
Thus, the theoretical calculation suggested a less distorted structure. 
If the two distorted MS WO$_3$ structures gathered and were aligned along the $c$-axis, along which the shortest W=O bonds were assumed to be headed, the distortion energy should be released.
Consequently, the formation of the well-ordered distorted MS WO$_3$ cluster shown in Fig. \ref{fig:polaron} should require a long time.  

The problem with the well-ordered MS WO$_3$ clusters is the charge balance, which should show a large negative charge.
To compensate for the negative charge, holes should combine with photoexcited electrons in the well-ordered MS WO$_3$ cluster. In this case, the recombination of photoexcited electrons and holes might be enhanced.  
If the spins of the hole and the photoexcited electron near the hole were  parallel or in a triplet state, the well-ordered distorted MS cluster should be stabilized and had a finite lifetime; however, we had little direct evidence supporting this scenario. 
If we could carry out large-scale DFT calculations of the MS excited state with a large cluster size, we could support our hypothesis for the formation of such a well-defined distorted MS WO$_3$ cluster.
Such calculations will be the topic of future work.  

\section{Conclusions}
\label{sec:conclusion}
Picosecond time-resolved W L$_3$-edge EXAFS spectroscopy was conducted to characterize the structural change in MS WO$_3$ in detail.  
We applied the CTS method to analyze the EXAFS data for the MS WO$_3$, where one of the shortest bonds was further shortened by 0.11 \AA \  from its original bond length, consistent with the structure proposed on the basis of W L$_1$-edge XANES spectroscopy.
This structural change in the MS WO$_3$ led the distorted MS structure, which should be stabilized, to form well-ordered distorted clusters in conjunction with the formation of the triplet state.

\section{Acknowledgement} 
The authors would like to express their gratitude to Prof. Masaaki Yoshida at Yamaguchi University for his technical advice and to the Japan Society for the Promotion of Science (JSPS) for their support through a Grant-in-Aid for Exploratory Research (No. 26620110), a Grant-in-Aid for Scientific Research (A) (Nos. 15H02173 and 20H00367), a Grant-in-Aid for JSPS Research Fellow (No. 15J07459), and a Grant for collaborative research in the Institute for Catalysis, Hokkaido University (No. 15A1004). 
The EXAFS experiments were carried out at beamline station BL-NW14A with the approval of PF-PAC (Prop. Nos. 2013G166, 2015G541, 2015G542).

See Supplemental Material at [URL will be inserted by publisher] for \ref{sec:XANES}Estimation of the ratio of photoexcitation;\ref{sec:Analysis} EXAFS data processing and the curve-fitting analysis;\ref{sec:RCF} Results of curve fitting;\ref{sec:CTS} Constrained thorough search; \ref{sec:GSWCTS} CTS analysis of the EXAFS for the WO$_3$ in the ground state ;\ref{sec:react} Surroundings-controlled W L$_3$-edge XANES measurements; \ref{sec:MO} Details of density functional theory (DFT) structure optimization for a single WO$_6$ complex in monoclinic and orthorhombic environments; \ref{sec:qmmm} Details of the QM/MM structural optimization for neutral, hole, and carrier states.

\bibliographystyle{elsarticle-num} 
\bibliography{askr2022bib}






\newpage




\renewcommand{\thefigure}{S\arabic{figure}}
\setcounter{figure}{0}
\renewcommand{\thetable}{S\arabic{table}}
\setcounter{table}{0}
\renewcommand{\theequation}{S\arabic{equation}}
\setcounter{equation}{0}
\renewcommand{\thesection}{ SI-\arabic{section}}
\setcounter{section}{0}
\setcounter{page}{1}

\date{\today}

\Large{Supporting Information} \\
\noindent
\ref{sec:XANES} Estimation of the ratio of photoexcitation\\
\ref{sec:Analysis} EXAFS data processing and curve fitting analysis\\
\ref{sec:RCF} Results of curve fitting \\
\ref{sec:CTS} Constrained thorough search \\ 
\ref{sec:GSWCTS} CTS analysis of the EXAFS data for WO$_3$ in the ground state \\
\ref{sec:react} Surroundings-controlled W L$_3$-edge XANES measurements \\
\ref{sec:MO} Details of the density functional theory (DFT) structure optimization for a single WO$_6$ complex in monoclinic and orthorhombic environments\\
\ref{sec:qmmm} Details about QM/MM structural optimization for neutral, hole, and carrier states\\
\newpage
\large
\section{ \label{sec:XANES} Estimation of the ratio of photoexcitation} 
\begin{figure}[hbpt]
    \centering
    \includegraphics[keepaspectratio,width=120mm]{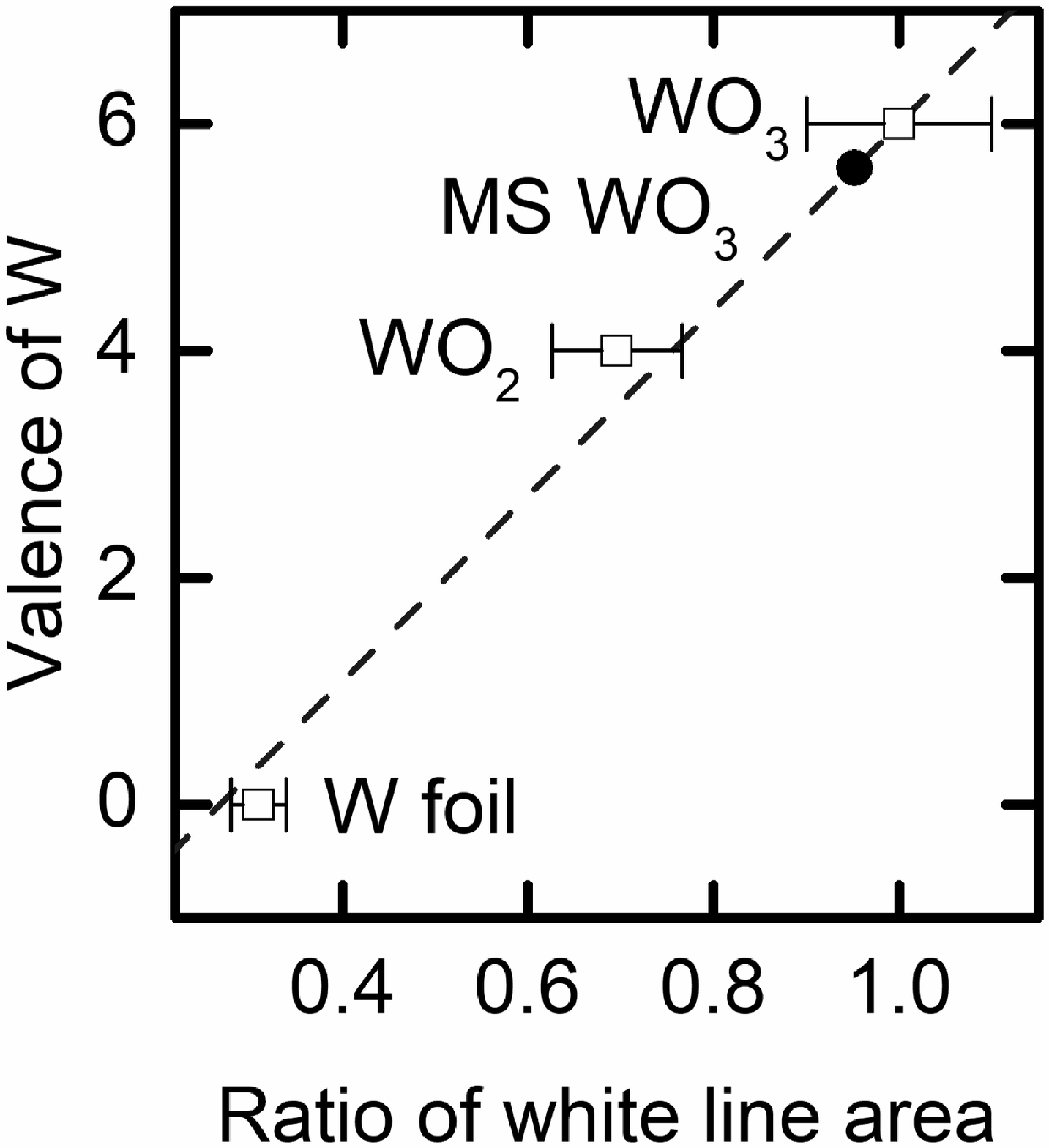}
     \vspace{-20truemm}
    \caption{(a) W L$_3$-edge XANES spectra of WO$_3$ before (black line) and 150 ps after  (red line) laser-pulse irradiation.  The blue dashed line shows the inverse tangent function for the baseline.  (b) The relationship between the ratio of the white-line area and the valence of W.  The black dot shows the result of the area ratio calculation from XANES spectra after laser irradiation. }
    \label{fig:XANES}
\end{figure}

\normalsize
The photoexcitation ratio was estimated from W L$_3$-edge XANES spectra.
The valence and conduction bands for WO$_3$ mainly consist of O 2$p$ orbitals and W 5$d$ orbitals, respectively. 
Thus, an electron is photoexcited from an O 2$p$ to a W 5$d$ orbital in the photoexcitation process of WO$_3$. 
In W L$_3$-edge XANES spectroscopy, an electron in a W 2$p_{3/2}$ orbital is excited to a W 5$d$ empty orbital at the edge region, which gives a strong absorption peak known as a white line. 
Therefore, the absorption coefficient in the white-line region decreased, as shown in Fig. \ref{fig:XANES}(a). 
The area ratio in the XANES spectrum before and after laser irradiation was 0.95. 
Fig. \ref{fig:XANES}(b) shows plots of the white-line area and the valence state of W for reference compounds W, WO$_2$, and WO$_3$.
From this plot, W$^{5+}$ or the MS WO$_3$ after photoirradiation was estimated as 40 $\pm$ 5\% \ for the total number of W. 

\newpage
\large
\section{EXAFS data processing and curve fitting analysis}
\normalsize
\label{sec:Analysis}
Data processing of EXAFS was carried out using REX2000 (RIGAKU Co).\cite{RN3616} 
The EXAFS oscillation $\chi\left(k\right)$ was derived from the absorption coefficient, $(\mu(E))\ $, by subtraction of the smooth background $\mu_B(E)$.
The $ \chi\left(k\right)$ is expressed as follows:

\begin{equation}
\begin{split}
\chi\left(k\right)&=\frac{\mu(E)-\mu_B(E)}{\mu_0(E)} \\ 
&={S_0}^2\sum_{i}{\frac{N_iF_i\left(k\right)}{kr_i^2}\exp{\left(-2 k^2{\sigma_i}^2\right)}\exp{\left(-2r/\lambda_i\right)} \sin(2kr_i+\ \phi_i\left(k\right))}    
\end{split}
\end{equation}
where $N_i$, $r_i$, and $\sigma_i^2$ are the coordination number, the bond length, and the Debye--Waller factor of the $i$th shell, respectively, and $F_i\left(k\right)$  and $\phi_i\left(k\right)$ are the backscattering amplitude and the phase shift, respectively, for the $i$th shell atom calculated using the FEFF8.2 code. 
Parameter $k$ is a wavenumber defined as 
\begin{equation}
k=\frac{\sqrt{2m(E-E_0-\Delta E_0)}}{\hbar} 
\end{equation}
where $m$ and $\hbar$ are the mass of an electron and the Dirac constant, respectively.
$E$, $E_0$, and $\Delta E_0$ are the incident X-ray energy, the absorption edge energy, and the correction factor for zero photoelectron kinetic energy, respectively.  
$S_0^2$ is the inelastic loss factor and is fixed to that for the reference compound Na$_2$WO$_4$.  
Curve fitting analysis was carried out using the LARCH package.\cite{RN6645}
The calculated and observed data were compared using R-factor.\cite{PhysRevB.48.9825}
\begin{equation}
   R-factor\ =\ \frac{\sum\left\{\chi_{obs}\left(k\right)-\chi_{cal}\left(k\right)\right\}^2}{\sum\left\{\chi_{obs}\left(k\right)\right\}^2} 
   \label{eq:RF}
\end{equation}

The amount of information was 
\begin{equation}
    M \approx \frac{2\Delta k \Delta r}{\pi} +2
\end{equation} 
where $\Delta k$ is the range of the Fourier transform and $\Delta r $ is the range of the inverse Fourier transform.  
\newpage
\large
\section{Results of curve fitting}
\label{sec:RCF}
\begin{figure}[bhpt]
    \centering
    \includegraphics[keepaspectratio,width=100 mm]{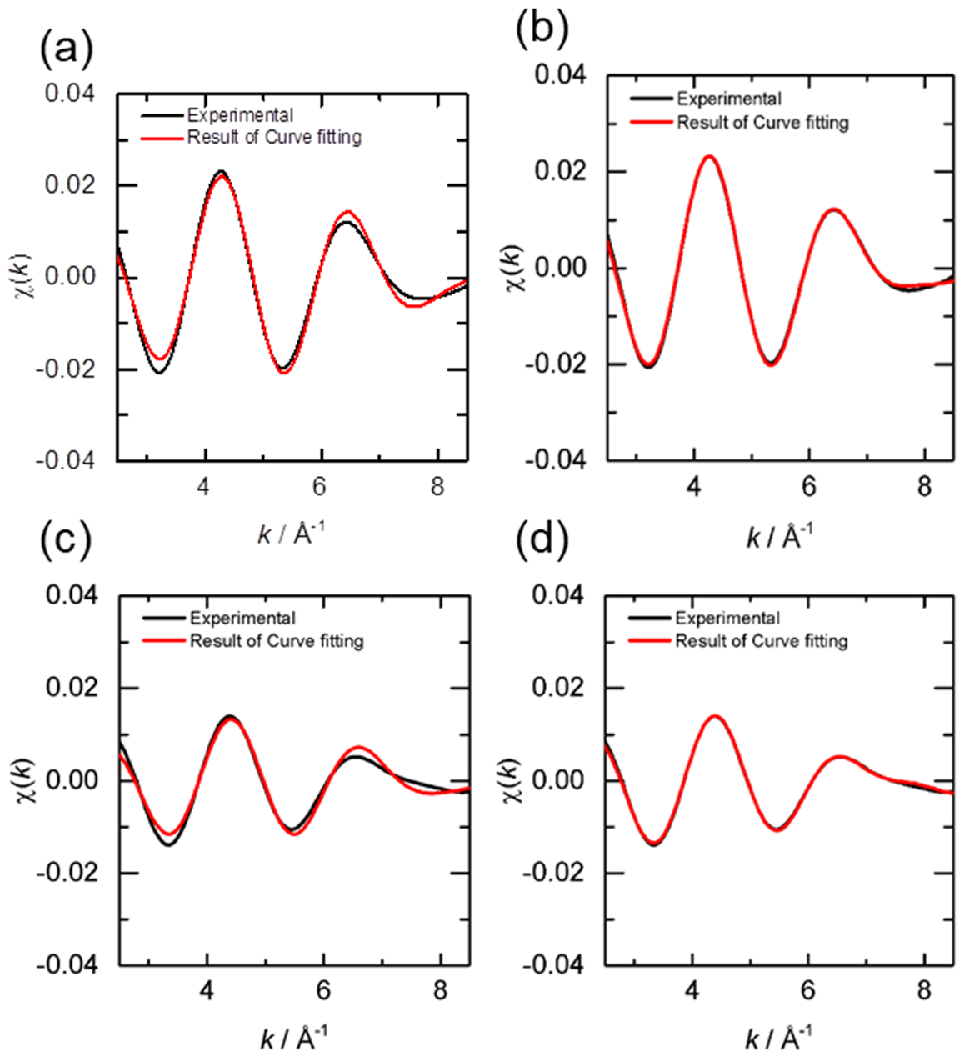}
    \caption{One-shell curve fitting results for GS (a,b) and MS (c,d). Debye--Waller factors are constrained by the bond length using eq. \ref{eq:DWANDR} (a,c) and free (b,d). The results are given in Table \ref{tab:CF1}.
  }
    \label{fig:CF}
\end{figure}

\normalsize
We carried out the curve fitting analysis on the WO$_3$ (ground state); the results are shown in Table \ref{tab:CF1}. 
The best fitting result is shown in Fig. \ref{fig:CF}.  
"Constrained" or "Free" indicates that the Debye--Waller factor was treated as a function of $r$ or was optimized as a fitting parameter, as described in the main text.  

\begin{table}[hbtp]
\caption{\label{tab:CF1}%
  Curve fitting results}
\scalebox{0.9}{
\begin{tabular}{l l l l l}
\hline
\textrm{Parameter }&
\textrm{Ground State}&&\textrm{Photoexcited State }& 
\\
\hline
&\textrm{Constrained} &\textrm{Free\ \ \ \ \ \ \ \ \ \ \ \  }&

\textrm{Constrained} &\textrm{Free\ \ \ \ \ \ \ \ \ \ \ \ \ }
\\
\hline
$N$&1.7&1.9&0.9&4.0\\
$r/$\AA \ &1.74&1.72&1.70&1.66\\
$\Delta$ E/eV &4.3&0.1&4.5&-4.1\\
$\sigma^2$/\AA$^2$&0.0014$^{*1}$&0.016$^{*2}$&0.0013$^{*1}$&0.028$^{*2}$\\
R-factor &0.01&0.001&0.03&0.001\\
\hline
\multicolumn{5}{l}{*1 Constrained case where  $\sigma^2$ was determined by Eq. \ref{eq:DWANDR}} \\
\multicolumn{5}{l}{*2 $\sigma^2$ was optimized. } 

\end{tabular}
}
\end{table}

\newpage
\large
 \section{Constrained thorough search}  
\begin{figure}[htbp]
    \centering
    \includegraphics[keepaspectratio,width=120 mm]{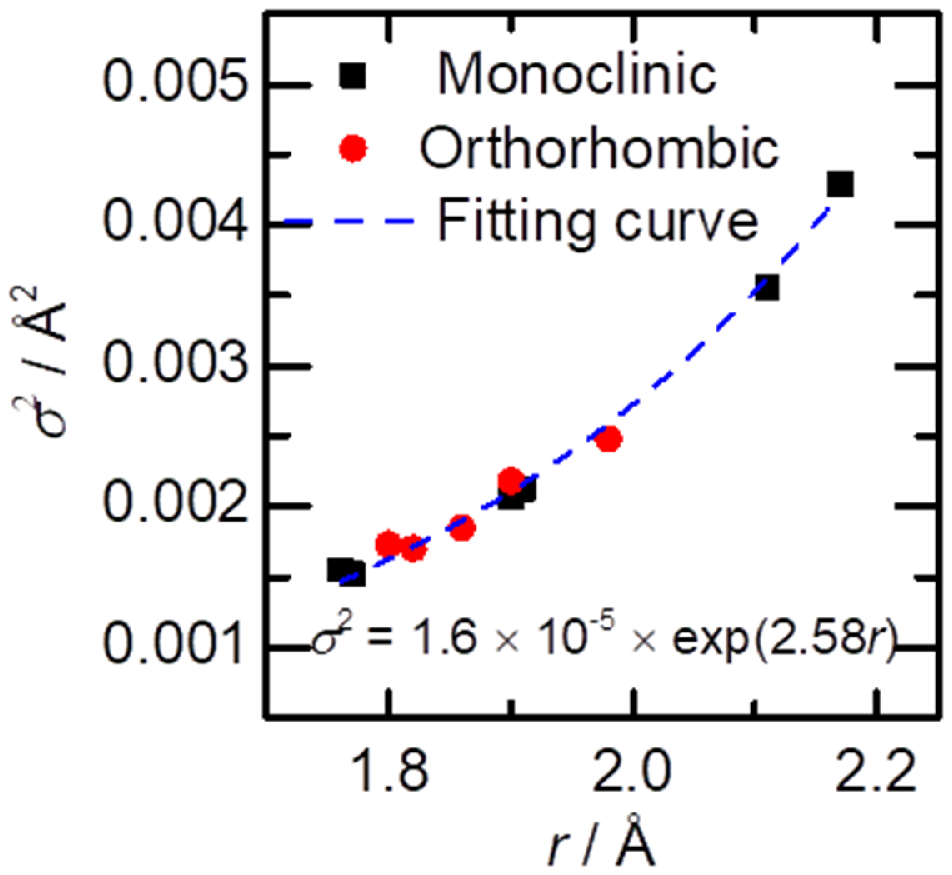}
\caption{ Results of the EM method for the W--O bonds of WO$_3$.  The dashed line shows the fitting results using the Debye--Waller factor ($\sigma^2$) versus the bond length ($r$). The bond lengths were obtained from the literature for monoclinic WO$_3$\cite{RN8155} and orthorhombic WO$_3$.\cite{RN8197}}
    \label{fig:DWTR}
\end{figure}

\begin{figure}[htbp]
    \centering
    \includegraphics[width=120mm]{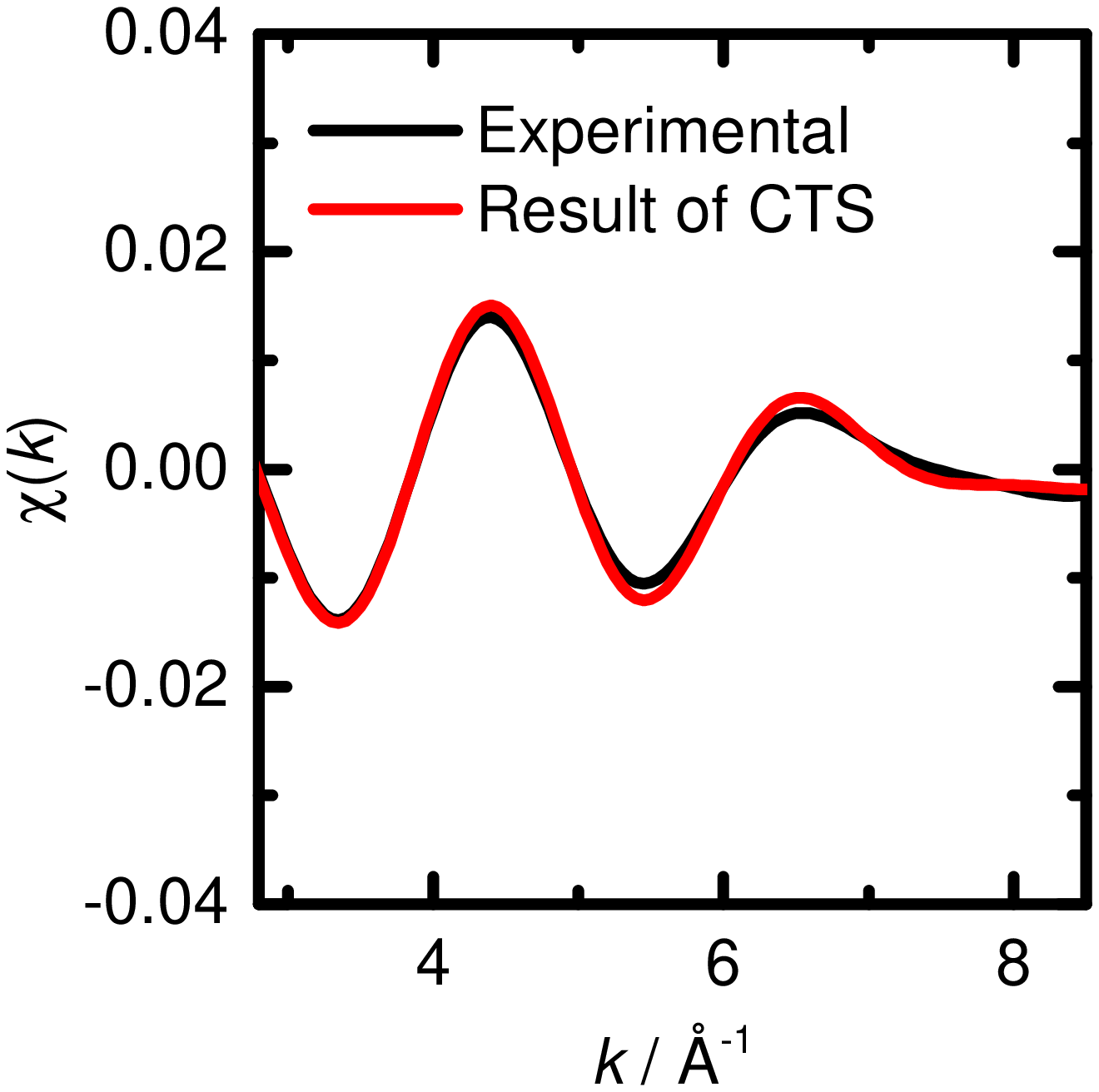}
   
    \caption{Comparison of $\chi_{MS}(k)$ between experimental data (black line) and calculated results based on the parameters determined by CTS (red line).  }
    \label{fig:TSFITTING}
\end{figure}

 \normalsize
 \label{sec:CTS}
A constrained thorough search (CTS) analysis was carried out using a custom-made Python program\cite{RN7005,RN7246} empowered by LARCH.\cite{RN6645}
In the CTS, we searched parameters within the $n$-dimensional parameter space. 
The value of $n$ was less than the amount of information, $M$. 
Thus, the constraints of several parameters were important. 
Because we determined the four W--O bond lengths separately, we fixed the coordination numbers for each bond as shown in Table \ref{tab:tableS1}. 
In this analysis, the $\Delta E_0$ and $S_0^2$ were fixed to those of the reference compound Na$_2$WO$_4$. 
The $\sigma^2$ values were estimated using the equation of motion (EM) method.\cite{RN8355}
The force constants for W--O bonds were obtained from the vibrational frequencies reported in the literature.\cite{RN7976} 
Because the force constants were strongly dependent on the bond lengths, \cite{RN1924} we  plotted the Debye--Waller factor as a function of the W--O distance,  as shown in Fig.\ref{fig:DWTR}. The relationship is approximated by the following equation:
  \begin{equation}
  \sigma ^2=A \times \exp(B \cdot r)     
  \label{eq:DWANDR}
  \end{equation}
where $A$ is $1.6\times 10^{-5}$ \AA$^2$ and $B$ is 2.58 \AA $^{-1}$.  
The surveyed range and steps for the CTS analysis are given in Table \ref{tab:tableS1}. 
We found all possible bond lengths in the R-factor $\leq$ 0.10 criterion based on the uniform prior probability principle.\cite{RN8640}  
We made histograms of the occurrence of each parameter, as shown in Fig. \ref{fig:HistoMS} or Fig.\ref{fig:histoGS}.
We fitted the histogram with a Gaussian function. 
Each average and the distribution corresponded to the bond length and its error. 
Fig. \ref{fig:TSFITTING} shows a comparison of $\chi(k)_{MS}$ and the calculated $\chi(k)$ using the structure parameters determined by the CTS method and reported in Table \ref{tab:CFR}.

\begin{table}[hbtp]
\caption{\label{tab:tableS1}%
  Survey ranges and steps of parameters for thorough search analysis.}
\scalebox{0.65}{
\begin{tabular}{llllllll}
\hline
\textrm{Parameter }&
\textrm{}&\textrm{Range / step }& A & B & Temp
\\
\hline
\textrm{Inelastic loss factor} &	$S_0^2$	&0.75 (Fixed) \\
\hline
\textrm{Correction for }\\ \textrm{the origin of kinetic energy } &	$\Delta E_0 $ /eV	17.0 (Fixed)& \\
\hline
Coordination number &$N_1$ & 1 (Fixed)\\
& $N_2$ & 1 (Fixed)\\
& $N_3$ & 2 (Fixed)\\
& $N_4$ & 2 (Fixed)\\
\hline
Bond length &$r_1$ & 1.50-1.90,0.01 step\\
&$r_2$ & 1.50-1.90,0.01 step\\
 &$r_3$ & 1.85-2.10,0.01 step\\
 &$r_4$ & 2.00-2.30 ,0.01 step\\
 \hline
 Debye--Waller factor & $\sigma^2$ / \AA$^2$ &	$A \times \exp(B \cdot r_i)$ & $1.6 \times 10^{-5}$& 2.58 \AA $^{-1}$  &   300 K \\
 \\
\hline
\end{tabular}
}
\end{table}
 
\begin{table}[hbtp]
\caption{\label{tab:CFR}
W--O bond length in ground-state WO$_3$, as determined from crystallographic data and CTS results\cite{RN8155}}

\scalebox{0.55}{
\begin{tabular}{l c c c c c c c  c c c}
\hline
$r'^*$ \ \ \ &Site1 & Distance  /\AA \ 	&$r''^*$\ \ \ &Site2 &Distance/ \AA \ 	&$r'''^*$\ \ \ & Average/ \AA \ 	&$r$\ \ \ &	unified Average/ \AA \ 	& 	CTS result / \AA\  \\
\hline
$r_1'$&W1-O6&	1.7434&	$r_1''$ & W2-O5&	1.7365& $r_1'''$ &  1.74 &	$r_1$ & 1.74&	1.77$\pm$0.08\\
$r_2'$ &W1-O3&	1.7776&	$r_2''$ &W2-O4&	1.7579& $r_2'''$ & 1.77 & $r_2$&	1.77&	1.77$\pm$0.08\\
$r_3'$& W1-O2&	1.8749&	$r_3''$ & W2-O2&	1.8899&	$r_3'''$ & 1.88 & $r_3$ & 1.90&	1.91 $\pm$0.04 
\\
$r_4'$ & W1-O1&	1.9209&$r_4''$ &	W2-O1&	1.8978&$r_4'''$ & 1.91 & &(Middle) &\\	
$r_5'$ & W1-O3&	2.0951&	$r_5''$ &W2-O4&	2.1009&	$r_5'''$&2.10 &$r_4$ &2.14& 2.17 $\pm$ 0.09 \\
$r_5'$ &W1-O5&	2.1845&	$r_5''$ &W2-O6& 2.1549&$r_5'''$ & 2.17 &&(Longest)		
\\
\hline
\multicolumn{11}{l}{$r',r'',r''' $indicate the W--O bond length around W1, W2, and their average, respectively.} 
\end{tabular}
}

\end{table}
\newpage

\newpage

\large

\newpage
\newpage
\large
\section{CTS analysis of the EXAFS data for WO$_3$ in the ground state }
\begin{figure}[htbp]
    \centering
    \includegraphics[keepaspectratio,width=120 mm]{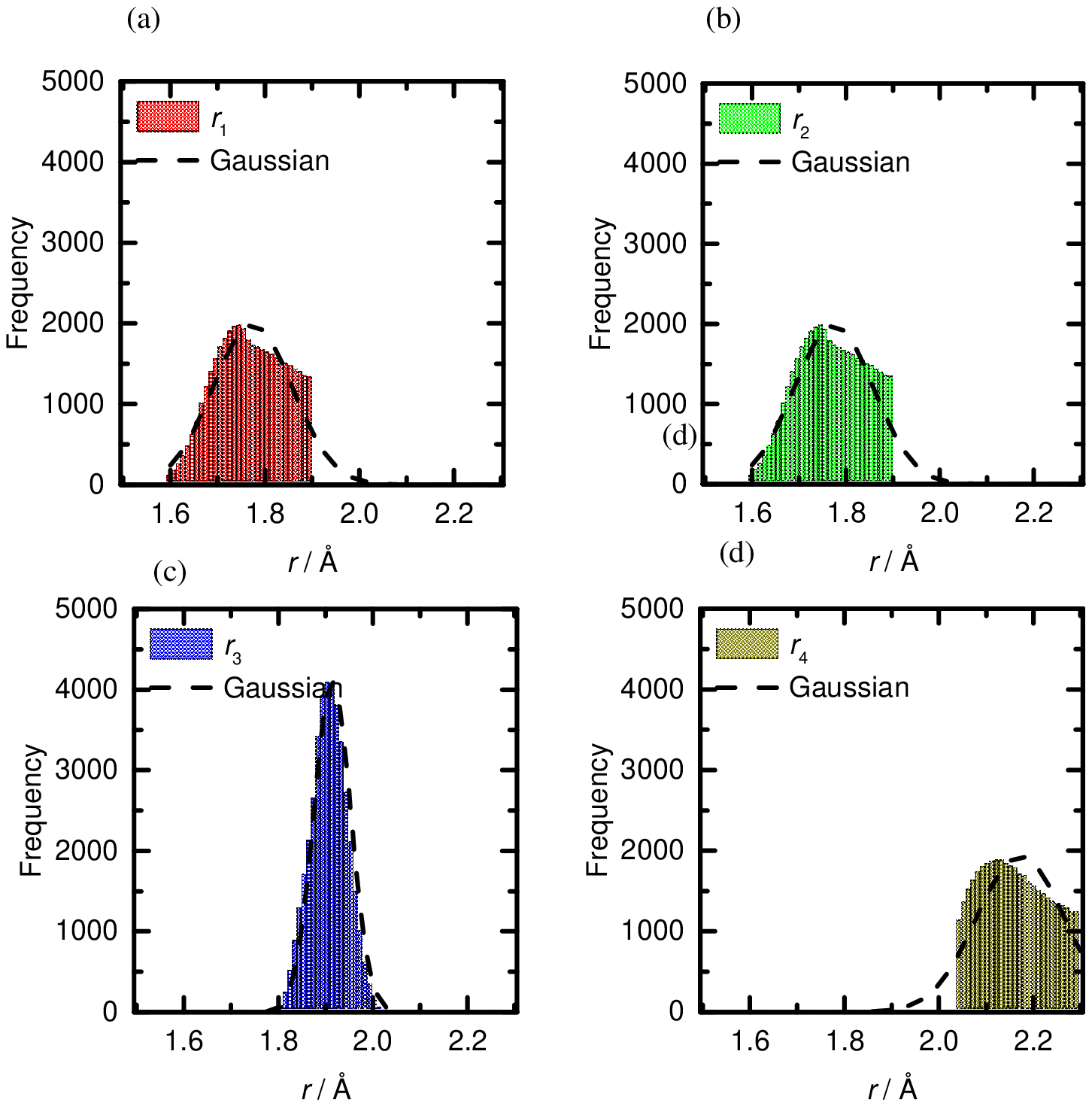}
    \caption{CTS histograms of bond lengths that satisfy the condition R-factor $\le$ 0.10.  Dashed lines show the fitting results obtained using a Gaussian function. Subfigures (a)--(d) correspond to the histograms of $r_1$, $r_2$, $r_3$, and $r_4$, respectively.}
    \label{fig:histoGS}
\end{figure}

\normalsize
\label{sec:GSWCTS}
Fig. \ref{fig:histoGS} shows CTS histograms of bond lengths that satisfy the condition R-factor $<$ 0.10. Dashed lines show the fitting results obtained using a Gaussian function.
The average value and standard deviation are the estimated result and its error, respectively. 
The structure parameters were determined and are given in Table \ref{tab:CFR}; they were found to be in good agreement with those reported in the literature.\cite{RN8155}   

\newpage
\large
\section{\label{sec:react} Surroundings-controlled W L$_3$-edge XANES measurements}
\begin{figure}[htbp]
    \centering
    \includegraphics[keepaspectratio,width=120 mm]{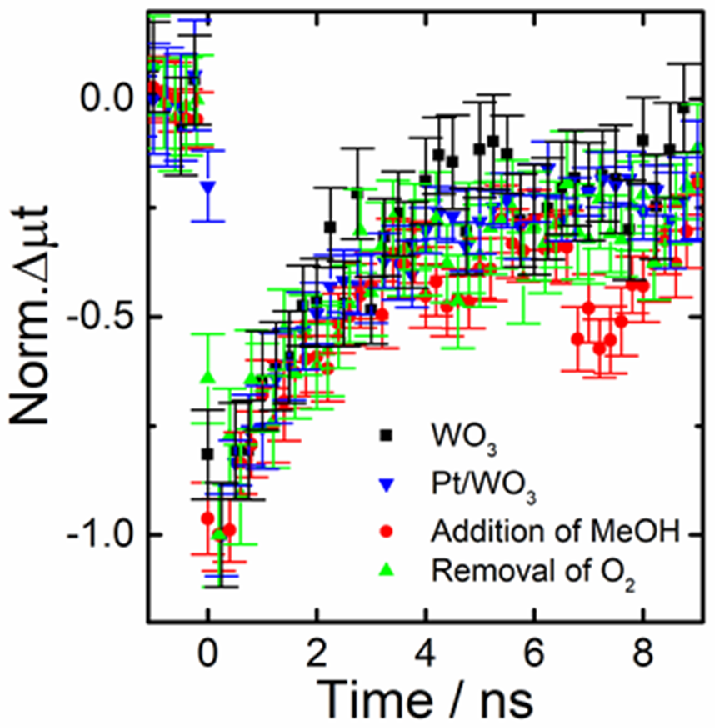}
    \caption{Time dependence of W L3-edge XANES spectra at 10,216 eV of WO$_3$ under surroundings-controlled conditions.   The height of the change was normalized at 200 ps after laser irradiation.}
    \label{fig:react}
\end{figure}

\normalsize
Picosecond time-resolved W L$_3$-edge XANES measurements were carried out under controlled surroundings. 
The conditions were as follows: (1) Pt particles loaded onto WO$_3$; Pt particles should enhance the charge transfer rate for excited electrons to the O$_2$ dissolved in water. (2) Removal of oxygen by nitrogen bubbling; the decrease in O$_2$ reduced the photoelectron consumption rate.  (3) Methanol addition at a concentration of 10 vol\%; the presence of methanol affected the lifetime of photo-generated holes.
The concentrations of the samples differed slightly, as shown in Table \ref{tab:REACT}.
However, the time dependence of the change in the W L$_3$-edge XANES spectra at 10,216 eV was the same, as shown in Fig. \ref{fig:react}.

\begin{table}[hbtp]
\caption{ 
Sample concentration and rate constant for surroundings-controlled W L3-edge XANES measurements}
\label{tab:REACT}
\scalebox{0.9}{
\begin{tabular}{l l l}
\hline
\textrm{Condition }&
\textrm{Concentration of WO$_3$ sample}&\textrm{Rate Constant }
\\
\hline
No additional treatment &	0.6 mmolL$^{-1}$ &	0.5 $\pm$ 0.1 /ns \\
Platinum loading &0.6 mmolL$^{-1}$ &	0.5 $\pm$ 0.1 /ns \\
Addition of methanol & 2.5 mmolL$^{-1}$ &	0.6 $\pm$ 0.1 /ns\\
Removal of oxygen &2.5 mmolL$^{-1}$ &	0.5 $\pm$ 0.1 /ns \\
\hline
\end{tabular}
}
\end{table}  

\newpage

\newpage
\section{\label{sec:MO} Details of the density functional theory (DFT) structure optimization for a single WO$_6$ complex in monoclinic and orthorhombic environments}
\begin{figure}[htbp]
    \centering
    \includegraphics[keepaspectratio,width=120 mm]{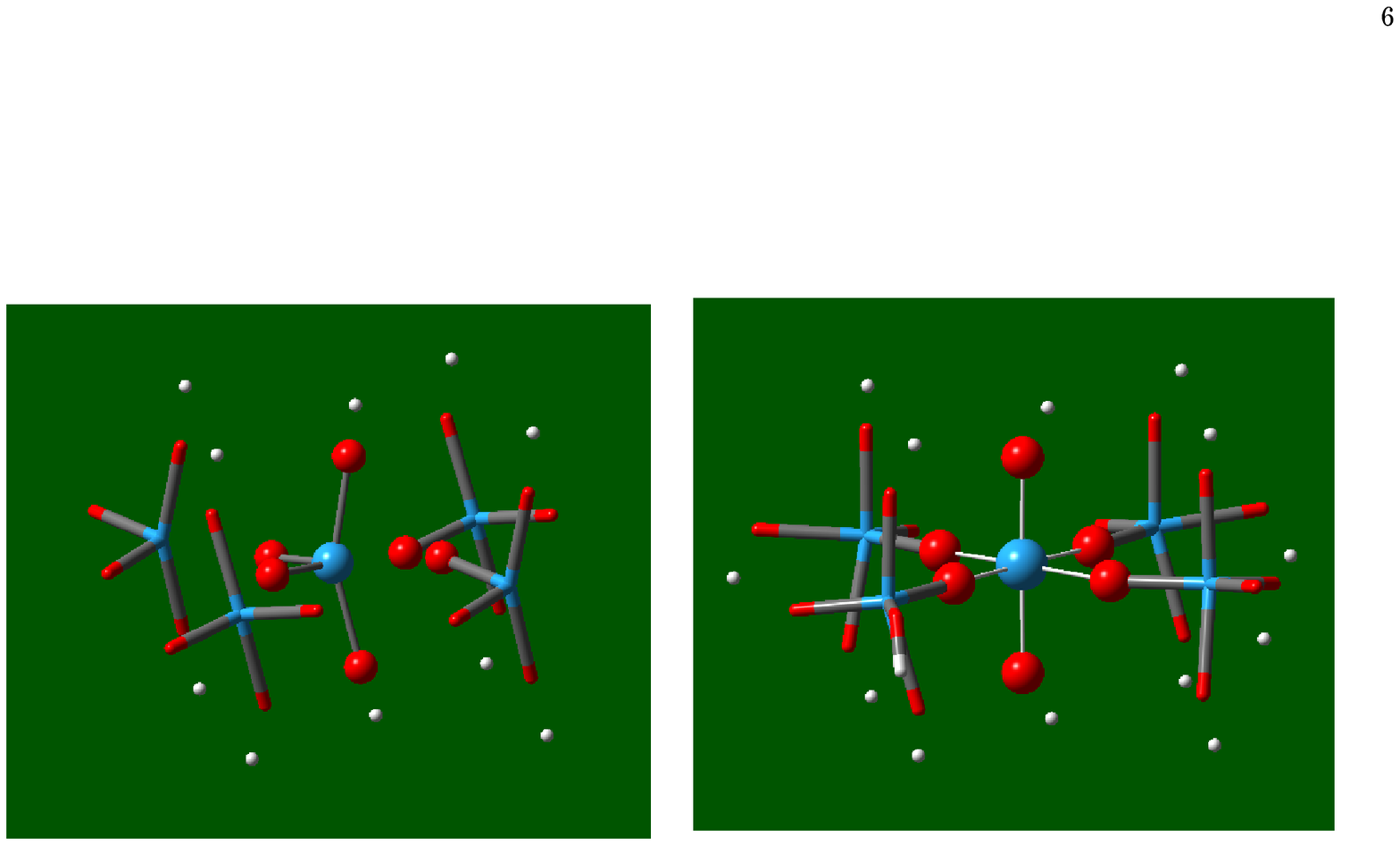}
     \vspace{-20 truemm}
    \caption{ Computational models for (a) the GS from monoclinic WO$_3$ and (b) the MS from orthorhombic WO$_3$. 
  }
    
    \label{fig:inistru}
\end{figure}

This computational part of the study was based on the following working hypothesis: Photoirradiation induces local excitation around a WO$_6$ unit. Very rapid structural relaxation occurs until the W complex is in the excited state. This structural change affects that around the W complex, and the structure transitions to the MS structure.\\
For the GS and MS structures, monoclinic \cite{RN8155} and orthorhombic WO$_3$ structures \cite{RN8197} were adopted. 
The computational models are shown in Fig. \ref{fig:inistru}.
The central WO$_6$ unit is surrounded by four W complexes in the same layer.
The W complexes in the upper and lower layers were omitted for simplicity.
 Instead, the open valency in the O atoms were capped by H atoms. For the structure optimization, only the central WO$_6$ unit was relaxed.
These models reproduced the original monoclinic and orthorhombic structures after the optimization process.
Electronic structure calculations were performed at the B3LYP/6-31G* level with Gaussian09.
For the excited-state calculation, time-dependent (TD) DFT was used.

\begin{figure}
[htbp]
    \centering
    \includegraphics[keepaspectratio,width=120 mm]{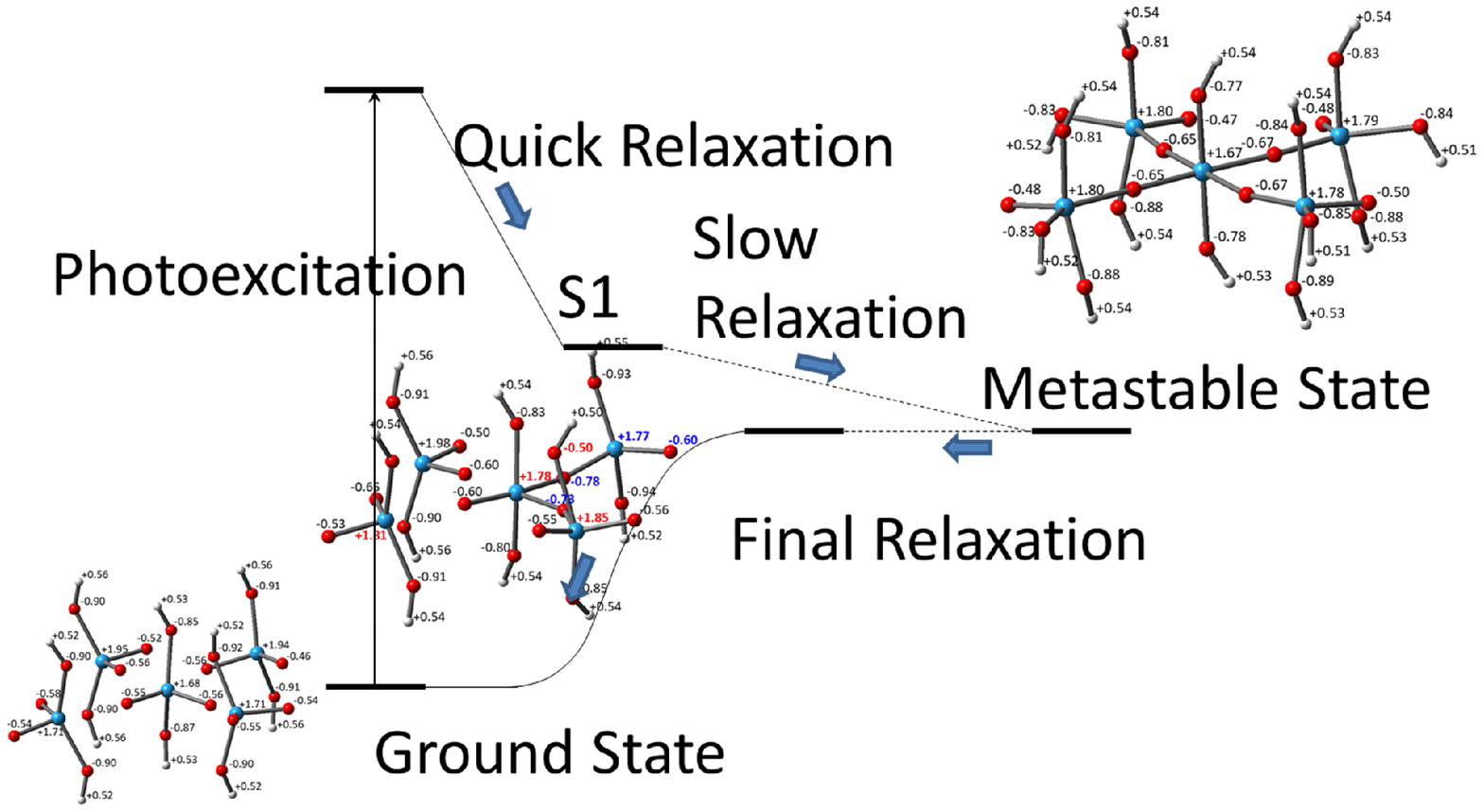}
    \caption{Energy diagram of photoabsorption process calculated by Gaussian 09.  The photoexcited species quickly relaxed to the S$_1$ state, and then to the metastable state (MS). Numerals in the figure show the natural population.  For the S$_1$ state, the numbers in blue+bold and red+bold denote that the decrease and increase, respectively, in the atomic change from the ground state is greater than 0.1.
  }
    
    \label{fig:Gaussian}
\end{figure}

 Fig. \ref{fig:Gaussian} shows an energy diagram of the excitation and relaxation processes. 
Table \ref{tab:GAUSSIAN}  shows the results.
The change in bond length in the MS state was rather symmetric; that is, shorter bonds were elongated, whereas longer ones were contracted.
The distortion was decreased even compared with that of the GS.
We carried out QM/MM analysis, as discussed in the next section.
We could not reproduce the more distorted structure in the MS WO$_3$, as observed in the W-L$_3$ edge EXAFS results.
This lack of reproducibility means that our working hypothesis does not match reality.
In particular, our orthorhombic structure assumption for the MS state is unrealistic.

\begin{table}[htbp]
    \centering
   \caption{Structural parameters of WO$_3$ in the ground state (GS), S$_1$ state, and MS state.}     \label{tab:GAUSSIAN} 

\scalebox{0.8}{
\begin{tabular}{l|ll|l|ll}\hline
Structural \ \  \ \  &\multicolumn{2}{c|}{\ \ MS(Monoclinic)\ \ }&S$_1$ state \ \ \ \ & \multicolumn{2}{c}{\ \ MS(Orthorhombic)\ \ }\\
\cline{2-3} 
\cline{4-4}
\cline{5-6}
parameters \ \ \ & Calc.( $\Delta$ )  \ \ \ \  &Exptl. \ \ &Calc.&Calc.($\Delta$) \ \  \ \  \ \  &Exptl. \ \  \\  
\hline 
W1-O2 &	1.96(+0.07)&	1.89&		1.93&		1.93(+0.07)&	1.86\\
W1-O1 &	1.97(+0.07)&	1.90&		1.90&		1.95(+0.09)&	1.86\\
W1-O3 &	1.75(-0.02)&	1.77&		1.84&		1.93(+0.11)&	1.82\\
W1-O6 &	1.75(0.00)&	1.75&		1.77&		1.89(+0.09)&	1.80\\
W1-O3 &	2.25(+0.16)&	2.09&		2.15&		1.96(+0.06)&	1.90\\
W1-O5 &	2.22(+0.07)&	2.15&		1.97&		1.96(-0.02)&	1.98\\
O3-W1-O6&	102.2(-0.4)&	102.6&		99.0&		92.0(-5.0)&	97.0\\
O3-W1-O3&	169.3(+4.6)&	164.7&		175.7&		178.5(-0.1)&	178.6\\
O5-W1-O6&	166.3(-4.6)&	170.9&		166.7&		176.7(+5.8)&	170.9\\
\hline

\end{tabular}
}
\end{table}

\newpage

%
\newpage

\section{Details about QM/MM structural optimization for neutral, hole, and carrier states \label{sec:qmmm} }

\par{Quantum mechanical/molecular mechanical (QM/MM) modeling is a method that embeds a DFT simulation within a lattice of point charges. These point charges are to simulate the long-range dielectric and structural properties of the system and are modeled using interatomic potentials.}

\par{The QM/MM method is particularly efficient for modeling systems with localized charged sites because the model allows for structural and dielectric changes to occur around the site while still maintaining the long-range structure and properties of the bulk material. In addition, the method avoids issues with periodic defects interacting with themselves, which is an issue that periodic boundary condition (PBC) modeling encounters, particularly with charged defects.}

\par{For this study, the FHI-aims software package \cite{RN8654} was used as the QM calculator, and the GULP software package \cite{RN8655} was used for the MM calculator. These two calculators were coordinated with the ChemShell software package \cite{RN8656}, which manages the overall QM/MM model.}

\par{QM/MM models were devised to investigate the structural changes associated with a localized electron hole, and separately a localized electron carrier within bulk WO\textsubscript{3}. These models were for the bulk material with the number of electrons within the QM-region altered to give the neutral material, as well as one missing an electron and one with an additional electron. These models were supported with neutral PBC models for the bulk material, as a basis from which to construct the QM/MM model but also to ensure accuracy and consistency, particularly for the orbital structure at the top of the valence band and at the bottom of the conduction band.}


\par{To generate the periodic model, the hybrid functional PBE0 was used, with the "light" basis sets of FHI-aims. The light basis sets were sufficient in this case because they give a sufficiently accurate representation of the geometric and electronic structure and because truly accurate values for energies were unnecessary. A $5\times 5\times 5$ $k$-grid was needed for meV convergence. }

\par{The QM/MM model was built from the relaxed structure of the periodic model following the procedure outlined in previous work \cite{RN8650} and was found to be reliable for studies of other metal oxides. The chosen QM-region was a $3\times 3\times 3$ grid of W\textsuperscript{6+} ions, and all the O\textsuperscript{2$-$} ions between them, totaling 81 atoms. }

\par{The potential energy functions used in the MM region were the Islam 1998 potentials \cite{RN8651}. These potentials involve Buckingham potentials between ions and a harmonic potential between anion cores and their electron shell. These simple potentials give the correct formation energies and dielectric constants for WO\textsubscript{3} but cannot replicate the complex structure of room-temperature monoclinic I ($\gamma$-WO$_3$) form. The partially covalent nature of WO\textsubscript{3} and the different lengths for its six W--O bonds due to pseudo Jahn--Teller distortions cannot be replicated with such simple potentials. As a consequence, all the MM nuclei needed to be fixed during energy minimization to prevent instability; if allowed to relax, the active MM region would shift to a cubic structure, creating spurious interactions at each of its boundaries. In addition, any ions in the QM region bound to multiple ions in the MM region were also fixed. }

\par{Such a strict fixing of ions is necessary for the model to work but, importantly, limits what it is capable of modeling. This model can only determine structural changes within a single WO\textsubscript{6} octahedron. If the correct structural change should be a particularly large-scale one, with multiple WO\textsubscript{6} octahedra distorting simultaneously, a substantially larger---and potentially computationally infeasible---calculation would be required. However, any understanding gained from knowing the electronic structure at the valence and conduction band boundaries, as well as the behavior of a fully localized system, can still be insightful.}


\par{In the excited state characterized by the XAFS studies, an electron is excited from the valence band into the conduction band. This excitation should result in the formation of an electron hole and a charge carrier. To investigate how this excitation process affects the structure and bond length, the neutral structure was energy-minimized in a periodic QM/MM model and then the structure with a hole was energy-minimized using QM/MM, then again for a carrier. This approach should reveal which (if either) state would influence bond lengths.}

\par{The HOMO is nonbonding with respect to any W--O bond; however, the LUMO is bonding with respect to the W-O5 bond, and antibonding with respect to the W-O2 bond.}

\par
\begin{figure}[hbt!]
\includegraphics[width=0.7
\textwidth]{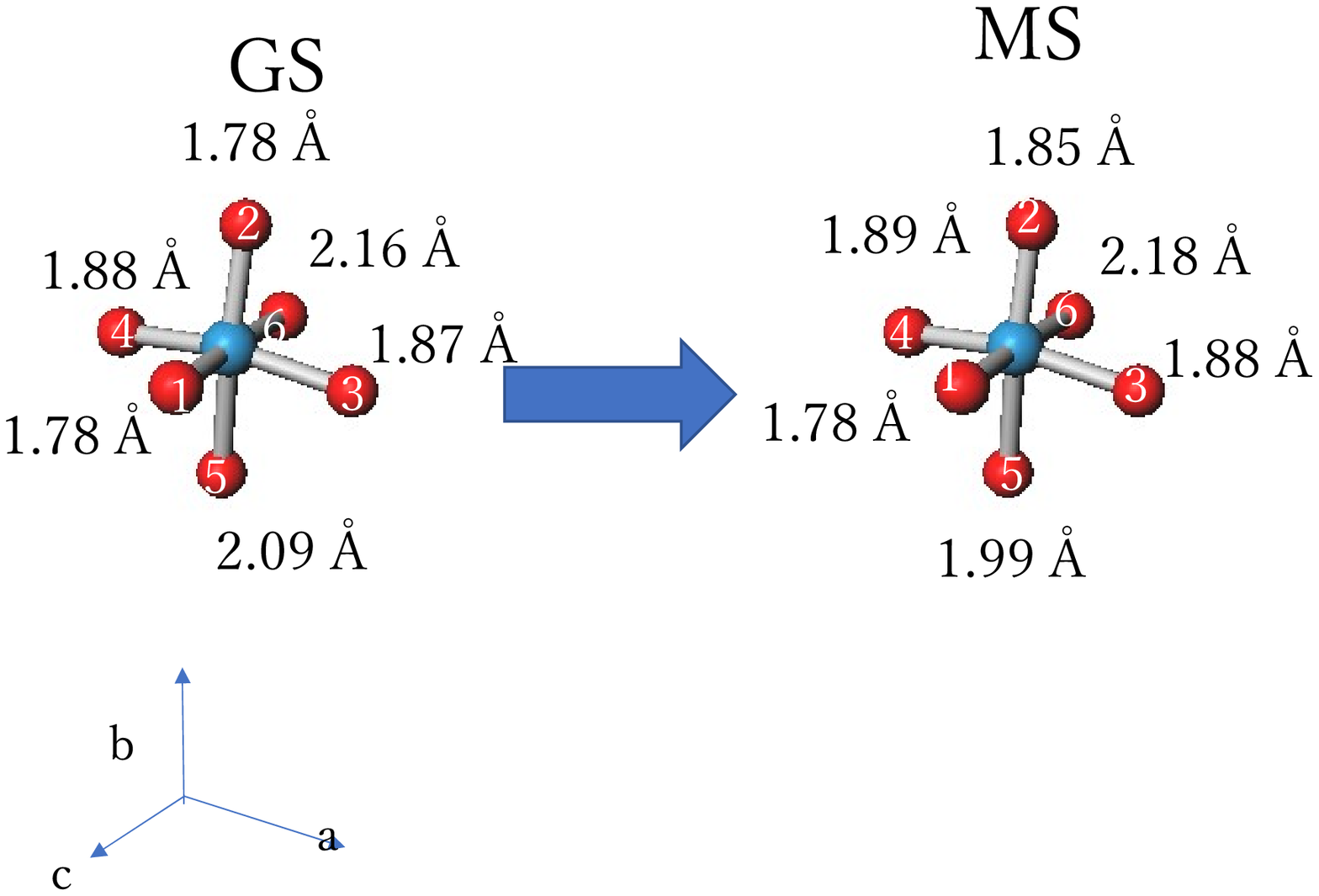} 
\caption{QM/MM results for structures of GS and MS WO$_3$}
\label{fig:WO3MSPPT}
\end{figure}

{Bond lengths for W-O1 through W-O6 were calculated for the relaxed neutral QM/MM models. The electron--hole and charge-carrier models varied from the neutral models by containing one fewer or one more electron in the QM region, respectively. The results are shown in Table \ref{fig:bond_lengths}, and are all expressed in angstroms. Figure \ref{fig:WO3MSPPT} shows the structures of the GS and MS WO$_3$. } 

\par{Reassuringly, the results for the neutral QM/MM models well match the experimental values. The maximum variance between models for any one bond length is approximately 0.04 \AA.}

\par{Between the neutral and hole QM/MM models, the bond lengths do not substantially change. This result was expected because removing an electron from an orbital that is neither bonding nor antibonding with respect to any W--O bond should not strongly affect any W--O bond length. The carrier QM/MM models, by contrast, showed a lengthening of the W-O2 bond and a shortening of the W-O5 bond compared with the neutral model. This result is expected because the LUMO orbital is bonding along the W-O2 bond and antibonding along the W-O5 bond. A spin-density analysis of the carrier model showed that the additional electron is mostly located on the central W (spin +0.52), with the rest located on the nearby W (spin +0.1 to +0.2).}


\begin{center}
\begin{table}
\caption{Values for W--O bond length for a range of models.}
\begin{tabular}{c c c c}
 \hline
 \ \ \ & \ \  QM/MM \ \ & \ \  QM/MM\ \  & \ \ QM/MM\ \  \\
 
 & Neutral & Hole & \underline{\textbf{Carrier}} \\
 \hline
 W-O1 (c)  & 1.78 & 1.77 & 1.78 \\
 W-O2 (b)  & \textbf{1.78} & 1.79 & \textbf{1.85} \\
 W-O3 (a) & 1.87 & 1.87 & 1.88 \\
 W-O4 (a) & 1.88 & 1.88 & 1.89 \\
 W-O5 (b) & \textbf{2.09} & 2.07 & \textbf{1.99} \\

 W-O6 (c) & 2.16 & 2.17 & 2.18 \\
 \hline
\end{tabular}
\label{fig:bond_lengths}
\end{table}
\end{center}

\end{document}